Short Paper*

# Enhancing E-Learning System through Learning Management System (LMS) Technologies: Reshape the Learner Experience

Cecilia P. Abaricia
Asian Institute of Computer Studies, Philippines
cecileabaricia@gmail.com
(corresponding author)

Manuel Luis C. Delos Santos
Asian Institute of Computer Studies, Philippines
ORCID: 0000-0002-6480-3377

## Abstract

*Purpose* – This paper aims to determine how the LMS Web portal application reshapes the learner experience through the developed E-Learning Management System using Data Mining Algorithm.

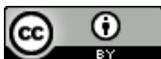

*Method* – The methodology that the researchers used is descriptive research involving the interpretation of the meaning or significance of what is described. Gather data from questionnaires, surveys, observations concerned with the study, and the chi-square formula for the statistical treatment of data.

*Results* – The findings of the study, the extent that LMS Web portal application reshapes the learner experience in terms of the following variables with the Average Weighted Mean (AWM): Flexible engagement of Learners in any device is highly satisfied; Personalize learning tracker is highly satisfied; Collaborating with the Learning Expert is highly satisfied; Provides user-friendly Teaching Tools is satisfied; Evident Learner Progress and Involvement and is satisfied.

*Conclusion* – In the final analysis, this E-Learning System can fit any educational needs as follows: chat, virtual classes, supportive resources for the students, individual/group monitoring, and assessment using LMS as maximum efficiency. Moreover, this platform can be used to deliver hybrid learning.

*Recommendations* – This LMS Web portal application is recommended for hybrid learning that helps to reshape the learner experience.

*Research Implications* – Through these platforms, learners provide an ease and flexible learning experience as self–paced learners. Students may work independently on their learning, updating on the students' performances and school announcements for the activities and events.

## INTRODUCTION

It is a great concept to use digital technology to increase learning. Web conferencing software such as Google Apps, Zoom, Open Educational Resources, and Learning Management Systems (LMS) are among the most often utilized technologies in Philippine education today. The learning management system (LMS) is a platform that educators may utilize to offer virtual and blended learning. It enables students to learn at any time and from any location.

Steindal et al. (2021) mentioned that during the onslaught of the pandemic covid-19, educational institutions worldwide have begun implementing learning management systems (LMSs) to centralize and manage learning resources, educational services, learning activities, and institutional information.

This paper aims to enhance the E-Learning System through LMS Technologies in reshaping their learner experience. Using this platform, the students can easily update the students' performances and school announcements for activities and events.

The goal of this study is to determine how LMS Web portal applications reshape the learner experience through the developed E-Learning Management System using Data Mining Algorithm.

Specifically, it sought to answer the following questions:

1. To what extent does the LMS Web portal application reshape the learner experience in terms of:

    1.1 Flexible engagement of Learners in any device
    1.2 Personalize learning tracker
    1.3 Collaborating with the Learning Expert
    1.4 Provides user-friendly Teaching Tools
    1.5 Evident Learner Progress and Involvement

2. Is there a significant difference in the perception of the two groups of respondents in determining how the LMS Web portal application reshapes the learner experience?

3. What are the challenges encountered by the teachers/school administrators in their student's learning?

4. How do the teachers/school administrators address the challenges encountered in their student's learning?

5. How does the LMS Web portal application reshape the learner experience based on the result of the study?

The scope of the study focused on the development of an LMS Web portal application using a data mining algorithm and how this application reshapes the learner experience in terms of flexible engagement of Learners in any device, personalize learning tracker, collaborating with the Learning Expert, provides user-friendly Teaching Tools and evident Learner progress and involvement. It aimed to determine the challenges encountered by the Asian Institute of Computer Studies teachers/school administrators in their students' learning. As a result of this study, the LMS Web portal application reshapes the learner experience. It involved school administrators, assistant school administrators, 10 teachers, and 30 students who served as respondents for the survey. The respondents were selected to answer the survey questionnaires using the purposive sampling technique.

The delimitation of the study focused on the lived experiences of the teachers/school administrators in their students' learning.

## LITERATURE REVIEW

### *Effect of Pandemic on Education*

In India, the online class mode was implemented to continue the education process during the Coronavirus outbreak that affected most sectors of society globally. In this mode, both teaching and learning experiences happen electronically via the internet (Salvaraj et al., 2021).

### *Learning Management System*

Based on the readings, a learning management system is software that helps to create, manage, organize, and deliver learning materials to students.

This system application simply implies according to its abbreviation: Learning provides up-to-date knowledge to the students; Management deals with organizing and managing online courses, teaching, and learning environment; and System pertains to interconnecting network.

A Learning Management System always consists of two parts: The admin interface is where the teachers can create, manage, and organize all the learning materials. It is enabled to generate reports (individual or group-based). This admin interface consists of settings and features to fully customize the learning materials. On the other, the user interface is where the students experience what an admin created. Students can access and participate in the created learning materials.

The importance of LMS help to retain information better through the learning materials uploaded where the students can access them anywhere, anytime. The students can set their own learning pace, which transforms their learning experiences. Second, it gives flexibility to busy work schedules and takes lesser time (What is an LMS? n.d.).

### *Learning Technology Trends*

LMS trends are always changing, which benefits both students and teachers. Learners are more engaged in modern systems, making job immersion more engaging. By planning experiences, students were more active participants in the process and gained access to learning that was tailored to their own needs. Now, LMSs provide automation of instruction as well as the monotonous record-keeping of the past. Engagement is an essential component of the learning process, and it is not a fad (Mardinger, n.d.).

### *Interactive Learning Technologies*

Online learning tools such as Zoom and Google Meet for teleconferencing (O'Neill, n.d.) seemed successful when their role was to complement the conventional classroom.

However, they have emerged as one of the primary methods of teaching strategy, and the possibility is that distance learning will remain even beyond the pandemic (Alkhaldi, 2020).

## *Distance Learning Challenges*

According to the readings, the actual difficulty is to deliver effective online learning with meaningful activities that keep students focused on their learning and performance goals.

Shorter online sessions and micro-lessons should be used instead of protracted academic-style video conferences to describe the learning situation. Teachers have uploaded video lessons that students may watch asynchronously at their leisure. The teachers also employed interactive, creative, and problem-solving activities.

Teachers may also believe it is necessary to formatively assess students' learning progress to give tailored feedback. Students can take online tests or play team activities.

Teachers are being urged to teach online during the Covid-19 global crisis to perform essential community duties. Using a learning management system (LMS) to promote school innovation (School Education Gateway, 2020).

## *Data Mining Algorithm*

Data mining is one of the technologies used to challenge how organizations implementing data warehousing technology can interpret the data and transform it into useful information and knowledge (Chen, et al., 2006).

## *Addressing Challenges*

In response to Porter's top-five list, Wesson provided the following roundup of resources and suggestions: Quality Instruction, Access for Students, Effective Communication, Social and Emotional Issues for Students, Teachers, Staff, and Administrators, and Economic Hardship for Students and Families are among the challenges (Top 5 Challenges, 2020).

## *Global Education Innovative Initiative*

Educating pupils in the twenty-first century entails nurturing competences in intrapersonal, interpersonal, and cognitive domains; establishing values, dispositions, and attitudes; and employing active, engaging, and empowering pedagogy (Reimers, 2020).

## METHODOLOGY

The research design used a descriptive research approach which is a basic research method that examines the situation (Williams, 2007) involving the interpretation of the meaning or significance of what is described. This methodology focuses more on the "what" of the research subject rather than the "why" of the research subject.

The research instruments used in this study were questionnaires and interviews. In data gathering, the survey questionnaire was one of the classical methods used to collect primary data (Dalati & Gomez, 2018) in this study regarding how the LMS Web portal application reshapes the learners' experience.

An interview for the teachers/school administrators regarding the challenges encountered and how to address the challenges encountered in their students' learning.

### *Statistical Treatment of Data*

To test the significant difference in the perception between two groups of respondents in determining how the LMS Web portal application reshapes the learner experience this study used the Chi-Square test method, developed by Karl Peterson in 1900 (Franke, et al., 2011) which can be used in evaluating any association between the rows and columns in a contingency table (Singhal & Rana, 2015).

### *Hardware and Software Components*

Figures 1 and 2 show the different components in terms of hardware and software in the development of the LMS web portal.

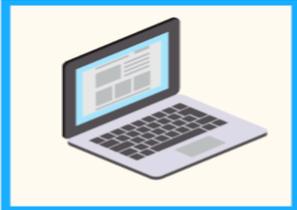

*Figure 1.* Hardware Components

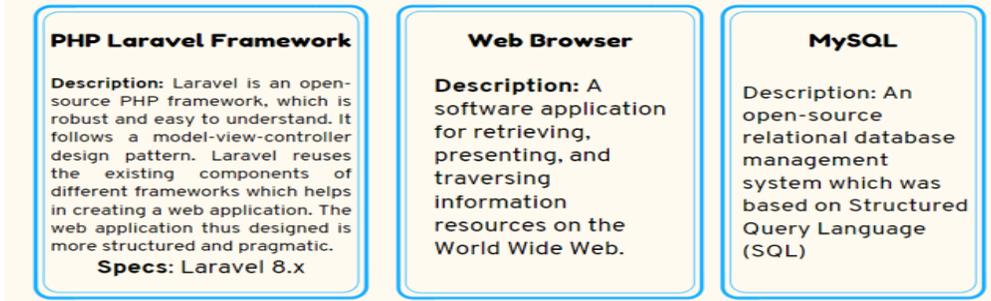
*Figure 2.* Software Components

Figures 3, 4, and 5 illustrate the project design of the LMS web portal which indicates the different roles and functionalities in terms of Admin, Teacher, and Student accounts.

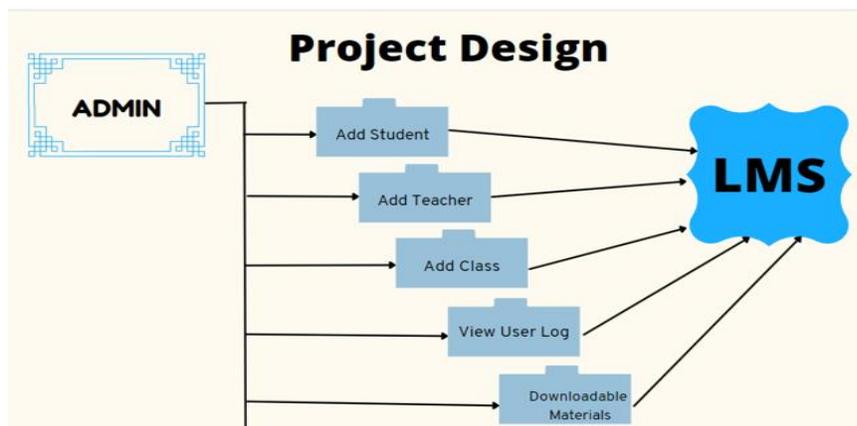
*Figure 3.* Project Design of the Admin Account

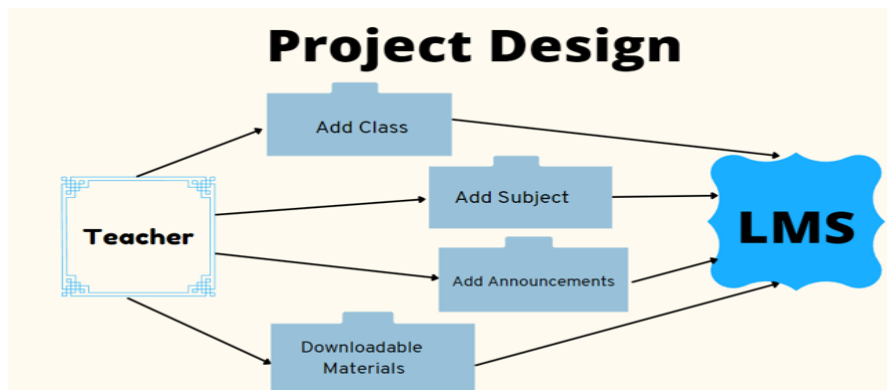
*Figure 4.* Project Design of the Teacher Account

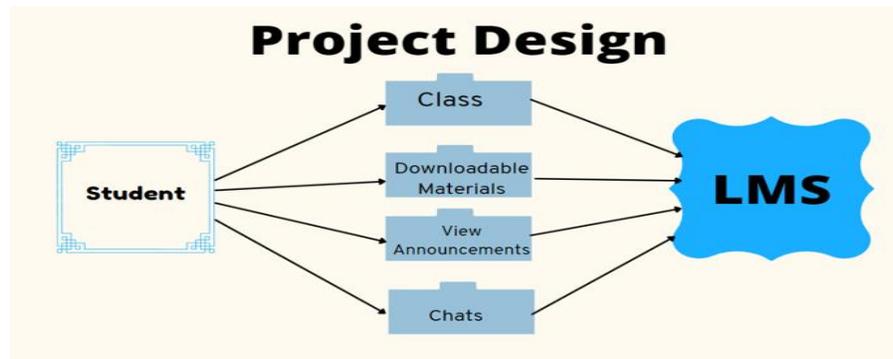

*Figure 5.* Project Design of the Student Account

Meanwhile, Figure 6 shows the integration of the algorithm which uses a data mining technique. The Mobile App Development Lifecycle is just a depiction of the process that consumes effort and resources (Vishnu, 2022) like the traditional Software Development Lifecycle (SDLC), but from the standpoint of a mobile device. Google's Android is an open-source, customizable mobile operating system designed for touchscreen smartphones. It is the most widely used operating system today. While it is most often seen on smartphones, it is also found on other smart devices like TVs and watches.

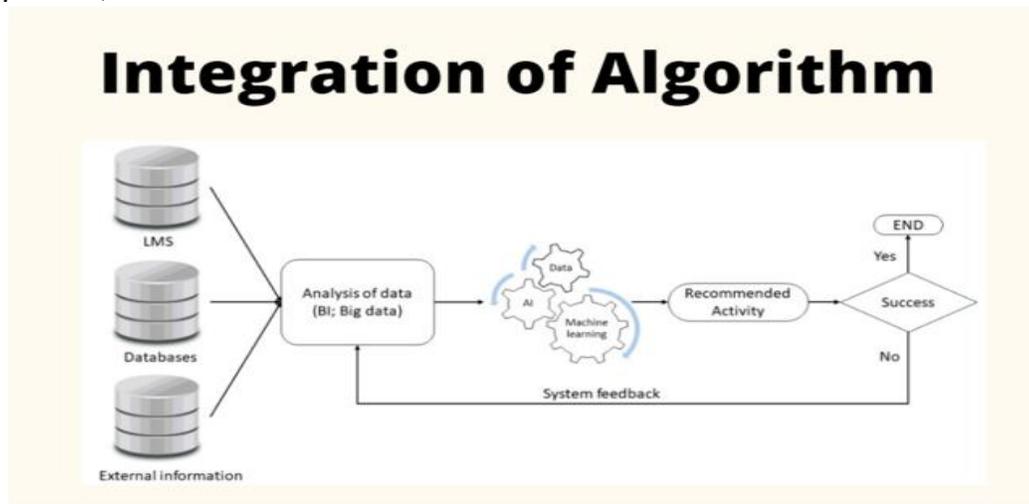

*Figure 6.* Integration of the Algorithm

The developers adhered to the eight stages of the Mobile Application Development Cycle. The first phase is the most essential since it is during this phase that the supporters set the framework for what comes afterward. Before going on to the next level, the proponents conduct extensive study and brainstorming.

This stage entails creating a comprehensive scope of work. The app's supporters went through mental prototyping and drew ideas on a whiteboard and even while creating a schematic. This is the first visual depiction of the ideas gathered in Phase 1 that assisted in identifying usability difficulties.

Understanding the graphics is not enough; supporters must also consider if the back-end infrastructure will support the App's functionality. To verify whether the App's concept is technically possible, supporters must have access to public data by simply sourcing public APIs to identify which platform is created for its platform (Android/iOS) as well as its format (tablet/smartphone).

A user cannot describe the touch experience unless he or she touches the App and observes how it functions and flows. To do this, the proponents must create a prototype and get the App experience into the hands of users as quickly as feasible.

The App must first be developed before it can be coded. The proponents build the interaction architecture of the design pieces during this phase. This is the stage in which a multi-step process and the outcomes have visual directives and blueprints that provide a visual representation of the ultimate output.

When the design is complete, the App may be built. Agile methodology is the greatest strategy for mobile app development since it allows you to adjust, add new features, and stay up with changing trends.

During this stage, the proponents must get some target consumers to test the App. Once it is complete, the App is ready to be submitted to the App stores for approval.

## RESULTS

As a result, the LMS Web portal application reshapes the learner experience. It provides educational resources such as lesson content resources, assessments, grades, and monitoring tools. Through LMS, the students can download learning materials, see their assignments, and attendance. It is easier for teachers and administrators to assess individualized activities based on student assessments. Besides, for the students, it can be helpful for them to better manage their learning.

### *Extent that LMS Technologies Reshape the Learner Experience*

Table 1 shows the extent that LMS Web portal application reshapes the learner experience in terms of the following variables with the Average Weighted Mean (AWM): Flexible engagement of Learners in any device is highly satisfied; Personalize learning tracker is highly satisfied; Collaborating with the Learning Expert is highly satisfied; Provides user-friendly Teaching Tools is satisfied; Evident Learner Progress and Involvement and is satisfied.

Table 1. Responses of Teachers/School Administrators and Students

| Variables | Administrators | | Students | | AWM | VI |
|---|---|---|---|---|---|---|
| | Mean | VI | Mean | VI | | |
| 1. Flexible engagement of learners in any device | 3.67 | HS | 3.53 | HS | 3.60 | HS |
| 2. Personalize learning tracker | 3.50 | S | 3.53 | HS | 3.52 | HS |
| 3. Collaborating with the Learning Expert | 3.67 | HS | 3.40 | S | 3.54 | HS |
| 4. Provide user-friendly Teaching Tools | 3.42 | S | 3.33 | S | 3.38 | S |
| 5. Evident Learner Progress and Involvement | 3.25 | S | 3.38 | S | 3.29 | S |
| AWM | 3.50 | S | 3.42 | S | 3.47 | S |

*Legend:* 4 – 3.51 - 4.00 **Highly Satisfied (HS)**  3 – 2.51 – 3.50 **Satisfied (S)**
2 – 1.51 – 2.50 **Slightly Satisfied (SS)**  1 – 1.00 – 1.50 **Not Satisfied (NS)**

## DISCUSSION

In addition, Table 1 presents the evaluation done by the teachers/school administrators with an Average Weighted Mean (AWM) is 3.50 with the verbal interpretation of satisfied. Then, it also shows the evaluation done by the students with the Average Weighted Mean (AWM) is 3.42 with the verbal interpretation of satisfied and the overall Average Weighted Mean (AWM) is 3.47 for the responses of teachers/school administrators and students that LMS with the verbal interpretation of satisfied.

Significant difference in the perception of the two groups of respondents in determining how LMS technologies reshape the learner experience is indicated in Table 2.

Table 2. Statistical Treatment of Data

| Computed Chi-square value |
|---|
| $\chi 2 = 0.13$ |

Ho: There is no significant difference in the perception of the two groups of respondents in determining how the LMS Web portal application reshapes the learner experience.

H1: There is a significant difference in the perception of the two groups of respondents in determining how the LMS Web portal application reshapes the learner experience.

$$df = (r-1)(c-1) = 3 \quad \text{Equation 1}$$

α = 5 % level of significance X2 = 7.815 (chi-square table) X2 (computed) < X2 (table)    *Equation 2*

Therefore, there is no significant difference in the perception of the two groups of respondents in determining how LMS technologies reshape the learner experience.

1. The challenges encountered by the teachers/school administrators in their students learning are as follows: Internet access problems for the students, Inadequate parental guidance for the students learning, decreasing motivation of students in studying due to certain circumstances, Financial difficulties for students' education and Effective communications among school staffs and teachers for quality education.

2. The teachers/school administrators address the challenges encountered in their students' learning by giving Learning Episodes for the topic discussed by the Teacher. Sending video lessons and instructional materials to the students for easy understanding of the lesson. Advising and motivating students to improve learning. Offering affordable quality education for the students. Disseminating clear and easy-to-understand instructions among the school employees.

## CONCLUSIONS AND RECOMMENDATIONS

In the final analysis, this e-learning system can fit any educational need as follows: chat, virtual classes, supportive resources for the students, individual or group monitoring, and assessment using the LMS for maximum efficiency. Moreover, this platform can be used to deliver hybrid learning.

This e-Learning system is recommended for online and face-to-face classes. It is a tool to help and reshape the learner experience.

## IMPLICATIONS

It is a great concept to use digital technologies to modify the learner's experience. Web conferencing software such as Google Apps, Zoom, Open Educational Resources, and Learning Management Systems are among the most often utilized technologies in Philippine education today (LMS). The learning management system (LMS) is a platform that educators may utilize to offer virtual and blended learning. It enables students to learn at any time and from any location.

## ACKNOWLEDGEMENT

The authors want to express their deep gratitude to the management of the Asian Institute of Computer Studies for sponsoring the presentation of this paper at the 1st International

Research Conference on Computer Engineering and Technology Education (IRCCETE) held at Fort Ilocandia Resort Hotel, Laoag City, Ilocos Norte, Philippines, last January 19-21, 2023. With sincere thanks to the Institute of Computer Engineers of the Philippines (ICpEP), the organizer of the event, and the publication of this paper to the International Journal of Computing Sciences Research (IJCSR). Above all, the Almighty God, for all the blessings, courage, and determination to finish this study.## DECLARATIONS

### Conflict of Interest

All authors declared that they have no conflicts of interest.

### Informed Consent

All participants were appropriately informed and voluntarily agreed to the terms with full consent before taking part in the conduct of the experiment.

### Ethics Approval

The AICS Research Ethics Committee duly approved this study after it conformed to the local and internationally accepted ethical guidelines.

## REFERENCES

Alkhaldi, N. (2020). *Evolution of Education: How Interactive Technologies Reshaped Education.* eLearning for Kids. https://elearningindustry.com/evolution-of-education-how-interactive-technologies-reshaped-learning

Chen, L. D., Sakaguchi, T. & Frolick, M. N. (2000). *Data Mining Methods, Applications, and Tools. Information Systems Management, 17*(1), 65-70, https://doi.org/10.1201/1078/43190.17.1.20000101/31216.9

Dalati, S. & Gomez, J. M. (2018). *Surveys and questionnaires. Modernizing the Academic Teaching and Research Environment: Methodologies and Cases in Business Research* (pp. 175-186). https://doi.org/10.1007/978-3-319-74173-4_10

Franke, T. M., Ho, T. & Christie, C. A. (2012). The Chi-Square Test: Often Used and Often *Misinterpreted. American Journal of Evaluation, 33*(3), 448-458. https://doi.org/10.1177/1098214011426594

Mardinger, R. (n.d). *Learning Technology Trends. What is an LMS? How to Choose the Right Learning Management System.* https://www.docebo.com/learning-network/blog/what-is-*learning*-management-system/

O'Neill, E. (n.d.). *The Best eLearning Tool You Need to Know.* LearnUpon Blog. https://www.learnupon.com/blog/best-elearning-tools/2077

## Author's Biography

**Cecilia P. Abaricia** is a graduate of BS in Computer Engineering from the Technological Institute of the Philippines. Also, has a Master's and Doctorate Degree in Technology from the Technological University of the Philippines. A certified professional Computer Engineer, a holder of TM 1 – Trainers Methodology, Computer System Servicing (CSS) NCII, Electronic Product Assembly and Servicing (EPAS) NCII TESDA Certified. A professor at the Tertiary level for almost 20 years, former Dean of the Engineering Department, with 10 years of Industry Experience, and a research/thesis adviser from 2004 – the present. She wrote a Teaching Guide in Chemistry 2 in TECHFACTORS, INC. and wrote Modules in BS Computer Science and BS Computer Engineering. An International Research Paper author and presenter, National Research Paper author and presenter, and published paper in the International Journal of Computing Science Research (IJCSR) and the Asian Citation Index ACI Indexed Journal.

**Manuel Luis C. Delos Santos** holds a bachelor's degree in the Bachelor of Science in Computer Science from the Divine Word College of Vigan, Ilocos Sur. A Master of Science in Computer Science and a Doctor in Information Technology at AMA University, Quezon City. He is a Civil Service Commission pen and paper test passer both at the Professional and Sub-Professional level, a curriculum developer and module writer in senior high school, tertiary education, and in a technical-vocational program. A Microsoft office specialist, former TESDA Assessor in CHS and CSS NCII, international and local research paper presenter, Scopus-indexed author, and co-author both in International and locally published research papers. Had been three times the best paper presentation awardee in 2018, 2019, and 2021, and two times the best poster presentation awardee in 2019 and 2021. An international journal committee member, peer reviewer, and research conference session chair. A research/thesis adviser and college professor from 2013 to the present. In addition, he is the current Dean of the Bachelor of Science in Computer Science degree program at the Asian Institute of Computer Studies (AICS) located in Quezon City, Philippines, the concurrent senior high school assistant program coordinator, and the head of the Center for Research and Development Unit of the AICSians Research Journal of AICS.